\documentclass{article}
\usepackage{graphicx}
\usepackage{amsmath}

\begin{document}

\noindent \textbf{Torque or no torque?! The resolution of the paradox using }

\noindent \textbf{4D geometric quantities with the explanation of the }

\noindent \textbf{Trouton-Noble experiment}\bigskip \bigskip

\qquad Tomislav Ivezi\'{c}

\qquad\textit{Ru%
\mbox
{\it{d}\hspace{-.15em}\rule[1.25ex]{.2em}{.04ex}\hspace{-.05em}}er Bo\v
{s}kovi\'{c} Institute, P.O.B. 180, 10002 Zagreb, Croatia}

\textit{\qquad ivezic@irb.hr\bigskip \medskip }

\noindent In this paper we have resolved the apparent paradox of different
mechanical equations for force and torque governing the motion of a charged
particle in different inertial frames. The same paradox arises in all usual
``explanations'' of the Trouton-Noble experiment. It is shown that the real
cause of the paradoxes is - \emph{the use of three dimensional (3D)
quantities}, e.g., $\mathbf{E}$, $\mathbf{B}$, $\mathbf{F}$, $\mathbf{L}$, $%
\mathbf{N}$, \emph{their transformations and equations with them.} Instead
of using 3D quantities we deal with \emph{4D geometric quantities, their
Lorentz transformations and equations with them}. In such treatment the
paradoxes do not appear. The explanation with \emph{4D geometric quantities }%
is in a complete agreement with the principle of relativity and with the
Trouton-Noble experiment. \bigskip \bigskip

\noindent \textbf{1. Introduction\bigskip }

\noindent In a recent paper Jackson [1] discussed the apparent paradox of
different mechanical equations for force and torque governing the motion of
a charged particle in different inertial frames. Two inertial frames $S$
(the laboratory frame) and $S^{\prime }$ (the moving frame) are considered. $%
S^{\prime }$ moves uniformly in the $+x$ direction with a speed $V=c\beta $.
A point charge $Q$ is fixed permanently at the origin in $S^{\prime }$. In $%
S^{\prime }$ a particle of charge $q$ and mass $m$ experiences only the
radially directed electric force caused by $Q$ at the origin. At time $%
t^{\prime }=0$ in $S^{\prime }$ the particle of charge $q$ is released at
rest with $r^{\prime }(0)=r_{0}^{\prime }$, see figure 1(a) in [1]. Such
initial conditions give that the particle has no angular momentum; it moves
radially outward without torque. Thus both the angular momentum $\mathbf{L}%
^{\prime }$ and the torque $\mathbf{N}^{\prime }$ are zero in $S^{\prime }$.
(Vectors in the three dimensional (3D) space will be designated in
bold-face.) In the laboratory frame $S$ the charge $Q$ is in uniform motion
and it produces \emph{both} an electric field $\mathbf{E}$ and a \emph{%
magnetic field} $\mathbf{B}$ that are given by equations (3a) and (3b)
respectively in [1]. The existence of the magnetic field $\mathbf{B}$ in $S$
is responsible for the existence of the 3D magnetic force $\mathbf{F}=q%
\mathbf{V}\times \mathbf{B}$ and this force provides a 3D torque on the
charged particle relative to the fixed origin in the laboratory $\mathbf{N}=%
\mathbf{x}\times \mathbf{F}$, see figure 1(b) in [1]. Consequently a
nonvanishing 3D angular momentum of the charged particle changes in time in $%
S$, $\mathbf{N}=d\mathbf{L}/dt$. Here we repeat Jackson's words [1] about
such result: ``How can there be a torque and so a time rate of change of
angular momentum in one inertial frame, but no angular momentum and no
torque in another? Is there a paradox? Some experienced readers will see
that there is no paradox - \emph{that is just the way things are}, ...'' (my
emphasis) Such reasoning is considered to be correct by many physicists.
However in the considered case \emph{the principle of relativity is violated}
and the ``explanation'' of the type ``\emph{that is just the way things are}%
'' does not remove the violation of the principle of relativity but only
accept that violation as something natural. We consider that such an
explanation as in [1] is not natural and not relativistically correct; the
paradox remained completely untouched in the approach from [1]. (In the
following the paradox examined in [1] will be called Jackson's paradox.)

In this paper it will be shown that - \emph{it is not the way things are},
but that there is a simple solution of the above problem which is in a
complete accordance with the principle of relativity. The real cause of the
paradox is - \emph{the use of 3D quantities}, e.g., $\mathbf{E}$, $\mathbf{B}
$, $\mathbf{F}$, $\mathbf{L}$, $\mathbf{N}$, \emph{their transformations and
equations with them.} \emph{The 3D quantities are considered as physical,
measurable quantities in the 4D spacetime.} Instead of using 3D quantities
we shall deal from the outset with \emph{4D geometric quantities, their
Lorentz transformations (LT) and equations with them}. In such treatment the
paradox does not appear and the principle of relativity is naturally
satisfied. \emph{It is considered in our approach that in the 4D spacetime
the physical reality, both theoretically and experimentally, is attributed
only to the 4D geometric quantities. }

The same paradox arises in all usual ``explanations'' of the Trouton-Noble
experiment. Here it will be shown that in the explanation with \emph{4D
geometric quantities }the Trouton-Noble paradox does not appear and such an
explanation is in a complete agreement with the principle of relativity and
with experiments.

In section 2 the standard transformations of the 3D $\mathbf{E}$ and $%
\mathbf{B}$ are quoted. In sections 3-5 different 4D geometric quantities
are introduced and discussed using geometric algebra formalism. This
includes the bivector field $F$, section 3, the 4D electric and magnetic
fields $E$ and $B$ (1-vectors) and the relations that connect $F$ with $E$
and $B$, section 4, then the 4D Lorentz force $K_{L}$ (1-vector), the
angular momentum $M$ (bivector) and the torque $N$ (bivector), section 5. In
section 6 we have presented the Lorentz transformations of the 4D $E$ and $B$
and of other multivectors. In section 7 the standard transformations of the
electric and magnetic field are derived and it is shown that they differ
from the Lorentz transformations of the 4D $E$ and $B$. The most important
sections are sections 8-8.4 and 9.2. The resolution of Jackson's paradox is
presented in four different ways in sections 8-8.4 using 4D geometric
quantities. In the same way the resolution of the Trouton-Noble paradox is
given in section 9.2. Finally section 10 refers to conclusions. \bigskip
\medskip

\noindent \textbf{2. Standard\ transformations of\ the\ 3D E and\ B\bigskip }

\noindent Both $\mathbf{E}$ and $\mathbf{B}$ given by (3a) and (3b) in [1]
can be also obtained using the relations that connect the 3D $\mathbf{E}$
and $\mathbf{B}$ in relatively moving inertial frames. In general they are
given by equation (11.149) from [2], which we repeat here
\begin{eqnarray}
\mathbf{E}^{\prime } &=&\gamma (\mathbf{E+\beta \times }c\mathbf{B})-(\gamma
^{2}/\gamma +1)\mathbf{\beta (\beta }\cdot \mathbf{E})  \notag \\
\mathbf{B}^{\prime } &=&\gamma (\mathbf{B+\beta \times E/}c)\mathbf{-(\gamma
^{2}/\gamma +}1\mathbf{)\mathbf{\beta (\beta }\cdot B).}  \label{Aeb}
\end{eqnarray}
The inverse transformations are found by interchanging primed and unprimed
quantities and putting $\mathbf{\beta \rightarrow -\beta }$. The
transformations (\ref{Aeb}) are derived by Lorentz [3], Einstein [4], and it
seems that according to [5] and [6] Poincar\'{e} was the first who gave a
mathematically valid derivation of the transformations of the 3D $\mathbf{E}$
and $\mathbf{B}$, see two fundamental Poincar\'{e}'s papers with notes by
Logunov [6]. According to such relations, e.g., the electric field $\mathbf{E%
}$ in one inertial frame is expressed by the mixture of $\mathbf{E}^{\prime
} $ and $\mathbf{B}^{\prime }$ from relatively moving inertial frame. They
are considered by almost all physicists to be the LT of the 3D $\mathbf{E}$
and $\mathbf{B}$, but for the reasons explained below, we shall call them
the standard transformations (ST), while the name LT will be reserved for
the LT of the 4D quantities.

In our case the relations (\ref{Aeb}), when written in components, become
\begin{eqnarray}
E_{x} &=&E_{x}^{\prime },\ E_{y}=\gamma (E_{y}^{\prime }+\beta
cB_{z}^{\prime }),\ E_{z}=\gamma (E_{z}^{\prime }-\beta cB_{y}^{\prime })
\notag \\
B_{x} &=&B_{x}^{\prime },\ B_{y}=\gamma (B_{y}^{\prime }-\beta E_{z}^{\prime
}/c),\ B_{z}=\gamma (B_{z}^{\prime }+\beta E_{y}^{\prime }/c).  \label{A1}
\end{eqnarray}
Denoting the event of the release of the particle as $A$ we write its
coordinates in $S^{\prime }$ as $A(ct_{A}^{\prime }=0,x_{A}^{\prime
},y_{A}^{\prime },z_{A}^{\prime }=0)$. Then the components of $\mathbf{E}%
^{\prime }(t_{A}^{\prime }=0)$ in $S^{\prime }$ are $E_{x}^{\prime
}(0)=kQx_{A}^{\prime }/r_{A}^{\prime 3}$, $E_{y}^{\prime
}(0)=kQy_{A}^{\prime }/r_{A}^{\prime 3}$, $E_{z}^{\prime }=0$, where $%
k=1/4\pi \varepsilon _{0}$, $x_{A}^{\prime }=r_{A}^{\prime }\cos \theta
_{A}^{\prime }$, $y_{A}^{\prime }=r_{A}^{\prime }\sin \theta _{A}^{\prime }$%
, $r_{A}^{\prime }$, $x_{A}^{\prime }$, $y_{A}^{\prime }$ and $\theta
_{A}^{\prime }$ are the values of $r^{\prime }$, $x^{\prime }$, $y^{\prime }$
and $\theta ^{\prime }$ at $t_{A}^{\prime }=0$, and also we have $\mathbf{B}%
^{\prime }=0$. (In [1] $r_{A}^{\prime }$ is denoted as $r_{0}^{\prime }$.)
The corresponding expressions for $\mathbf{E}$ and $\mathbf{B}$ in $S$ are
obtained in all usual approaches to electromagnetism by the use of the ST (%
\ref{A1}). They are
\begin{eqnarray}
E_{x} &=&kQ\gamma (x_{A}-\beta ct_{A})/\varsigma ^{3},\ E_{y}=kQ\gamma
y_{A}/\varsigma ^{3},\ E_{z}=0  \notag \\
B_{x} &=&B_{y}=0,\ B_{z}=\beta E_{y}/c  \label{AE}
\end{eqnarray}
where $\varsigma =[\gamma ^{2}(x_{A}-\beta ct_{A})^{2}+y_{A}^{2}]^{1/2}$ and
the LT of the coordinates of the event $A(0,x_{A}^{\prime },y_{A}^{\prime
},0)$ are employed, $ct_{A}^{\prime }=0=\gamma (ct_{A}-\beta x_{A}),$ $%
x_{A}^{\prime }=\gamma (x_{A}-\beta ct_{A}),$ $y_{A}^{\prime }=y_{A},$ $%
z_{A}^{\prime }=z_{A}=0$. Jackson [1] assumed that not only $t_{A}^{\prime
}=0$ than also $t_{A}=0$. With such an assumption the relations (\ref{AE})
become equations (3a) and (3b) in [1] but, in fact, it is not correct in
this case to take that $t_{A}=0$ as well. Namely the event of the
coincidence of the origins of $S^{\prime }$ and $S$, let it be $O$, has the
coordinates $O(t_{O}^{\prime }=0,0,0,0)$ in $S^{\prime }$ and $%
O(t_{O}=0,0,0,0)$ in $S$. Thus the events $O$ and $A$ are simultaneous in $%
S^{\prime }$, $t_{O}^{\prime }=t_{A}^{\prime }=0$ and they cannot be
simultaneous at the same time in $S$, i.e., $t_{A}$ must be $\neq 0$.
However we are not interesting in it since only what is important here is
the appearance of $B_{z}\neq 0$ in $S$. This leads to $d\mathbf{L}/dt$ and $%
\mathbf{N}$ different from zero in $S$ and thus to the violation of the
principle of relativity in the laboratory frame $S$. \bigskip \medskip

\noindent \textbf{3. The\ electromagnetic field }$F$\textbf{\bigskip }

\noindent Now consider the same problem using geometric 4D quantities. This
investigation will be done in the geometric algebra formalism which is
presented in [7-11]. Physical quantities will be represented by geometric 4D
quantities, multivectors that are defined without reference frames, i.e., as
absolute quantities (AQs) or, when some basis has been introduced, they are
represented as 4D coordinate-based geometric quantities (CBGQs) comprising
both components and a basis. Usually [7-11] one introduces the standard
basis. The generators of the spacetime algebra are taken to be four basis
vectors $\left\{ \gamma _{\mu }\right\} ,\mu =0,...3$ (the standard basis)
satisfying $\gamma _{\mu }\cdot \gamma _{\nu }=\eta _{\mu \nu }=diag(+---).$
This basis is a right-handed orthonormal frame of vectors in the Minkowski
spacetime $M^{4}$ with $\gamma _{0}$ in the forward light cone. The $\gamma
_{k}$ ($k=1,2,3$) are spacelike vectors. The basis vectors $\gamma _{\mu }$
generate by multiplication a complete basis for the spacetime algebra: $%
1,\gamma _{\mu },\gamma _{\mu }\wedge \gamma _{\nu },\gamma _{\mu }\gamma
_{5,}\gamma _{5}$ ($16$ independent elements). $\gamma _{5}$ is the
pseudoscalar for the frame $\left\{ \gamma _{\mu }\right\} $, $\gamma
_{5}=\gamma _{0}\wedge \gamma _{1}\wedge \gamma _{2}\wedge \gamma _{3}$.

It is worth noting that the standard basis corresponds, in fact, to the
specific system of coordinates that we call Einstein's system of
coordinates. In Einstein's system of coordinates the standard,
i.e.,Einstein's synchronization [4] of distant clocks and Cartesian space
coordinates $x^{i}$ are used in the chosen inertial frame. However \emph{%
different systems of coordinates of an inertial frame are allowed and they
are all equivalent in the description of physical phenomena. }For example,
in [12,13] and in the second and the third paper in [14], two very
different, but physically completely equivalent systems of coordinates,
Einstein's system of coordinates and the system of coordinates with a
nonstandard synchronization, the everyday (radio) (``r'') synchronization,
are exposed and exploited throughout the paper. For the sake of brevity and
of clearness of the whole exposition, we shall mainly work with the standard
basis $\left\{ \gamma _{\mu }\right\} $, but remembering that the approach
with 4D quantities that are defined without reference frames holds for any
choice of basis.

Note that our living arena is the 4D spacetime in which, according to our
opinion, physical reality, both theoretically and \emph{experimentally}, is
attributed only to geometric 4D quantities, AQs or CBGQs, and to physical
laws expressed by such geometric 4D quantities. When physical laws are
written with 4D AQs or 4D CBGQs then there is no room for the preference of
any synchronization, standard or nonstandard, or, better to say, of any
system of coordinates even in an inertial frame. This is examined in a
geometric approach to special relativity (SR), i.e., the invariant SR, which
is developed in [12-19] and compared with experiments in [14] and [17-19].
(The name invariant SR comes from the fact that such geometric approach to
SR exclusively deals with AQs or with the corresponding CBGQs and every CBGQ
is invariant upon the passive LT; the components transform by the LT and the
basis by the inverse LT leaving the whole CBGQ unchanged. This will be
explained in section 6.) In addition we remark that the usual covariant
formalism does not work with geometric quantities but only with components
(numbers) taken usually in the $\left\{ \gamma _{\mu }\right\} $ basis; the
basis is only implicit not explicit in the covariant formalism.

Although we shall utilize the geometric algebra formalism in a manner very
similar to that one in the above mentioned references [7-11], our results in
the electromagnetic field theory markedly differ from all previous results
including [1-11]. These results are already published in the tensor
formalism [12,13,16] (with tensors as AQs or equivalently as CBGQs) and also
presented as e-prints [15], [17-19] both in tensor and geometric algebra
formalisms. (In [12] and again in [13] it is found in a manifestly covariant
way that there is, contrary to the generally accepted opinion, a
second-order electric field outside stationary superconductor with steady
current.) It is important to note that these new results are completely in
agreement with the principle of relativity and with experiments that test SR
as can be clearly seen, e.g., from [17-19] and particularly [14].

First let us write the bivector field $F(x)$ (or, we shall also call it the
electromagnetic field $F(x)$) for a charge $Q$ with constant velocity $u_{Q}$
(1-vector), see, e.g., [10] equation (7.94) or [11] equation (26), or the
discussion in [19] section IV.B,
\begin{equation}
F(x)=kQ(x\wedge (u_{Q}/c))/\left| x\wedge (u_{Q}/c)\right| ^{3}.  \label{cvf}
\end{equation}
In (\ref{cvf}) $F(x)$ is written as an AQ, i.e., it is defined without
reference frames. For the charge $Q$ at rest, $u_{Q}/c=\gamma _{0}$, whence
\begin{equation}
F_{(0)}(x)=kQ(x\wedge \gamma _{0})/\left| x\wedge \gamma _{0}\right| ^{3}.
\label{cv2}
\end{equation}

All AQs in equations (\ref{cvf}) and (\ref{cv2}) can be written as CBGQs in
some basis. We shall write them in the standard basis $\{\gamma _{\mu }\}$.
In the $\{\gamma _{\mu }\}$ basis $x=x^{\mu }\gamma _{\mu }$, $%
u_{Q}=u_{Q}^{\mu }\gamma _{\mu }$, $F=(1/2)F^{\alpha \beta }\gamma _{\alpha
}\wedge \gamma _{\beta }$ (the basis components $F^{\alpha \beta }$ are
determined as $F^{\alpha \beta }=\gamma ^{\beta }\cdot (\gamma ^{\alpha
}\cdot F)=(\gamma ^{\beta }\wedge \gamma ^{\alpha })\cdot F$). \bigskip
\medskip

\noindent \textbf{4. The\ relations\ that\ connect\ }$F$\textbf{\ with\ 4D\ }%
$E$\textbf{\ and\ }$B$\textbf{\bigskip }

\noindent From the given $F$ one can construct electric and magnetic fields
represented by different algebraic objects, e.g., 1-vectors or bivectors.
Instead of using the spacetime split and the bivectors (relative vectors and
relative bivectors) for the representation of the electric and magnetic
fields as in [7-11], we shall make an analogy with the tensor formalism [20]
and represent the electric and magnetic fields by 1-vectors $E$ and $B$ that
are defined without reference frames, i.e., as AQs. Such representation with
1-vectors $E$ and $B$ and their real and complex combination is examined in,
e.g., [15] and also in [17-19]. (The formulations of the classical
electromagnetism in terms of the 4-vectors (components not geometric
quantities) of the electric $E^{\alpha }$ and magnetic $B^{\alpha }$ fields
are presented in [21-23] in the usual covariant tensor formalism. In [23]
the relativistically correct definition of the electromagnetic 4-momentum
with $E^{\alpha }$ and $B^{\alpha }$ is presented and used to resolve the
famous ``4/3'' factor appearing in the problem of the electromagnetic mass
of the classical electron.) The electric and magnetic fields defined without
reference frames, i.e., \emph{independent of the chosen reference frame and
of the chosen system of coordinates in it}, thus as AQs, are given as
\begin{align}
F& =(1/c)E\wedge v+(IB)\cdot v  \notag \\
E& =(1/c)F\cdot v,\quad B=-(1/c^{2})I(F\wedge v)  \label{itf}
\end{align}
where $I$ is the unit pseudoscalar. ($I$ is defined algebraically without
introducing any reference frame, as in [24], section 1.2.) The velocity $v$
and all other quantities entering into the relations (\ref{itf}) are AQs.
That velocity $v$ characterizes some general observer. We can say, as in
tensor formalism [20], that $v$ is the velocity (1-vector) of a family of
observers who measures $E$ and $B$ fields. Of course \emph{the relations for}
$E$ \emph{and }$B$, equations (\ref{itf}),\emph{\ are coordinate-free
relations and thus they hold for any observer.} The relations (\ref{itf})
are \emph{manifestly Lorentz invariant equations}. Note that $E\cdot
v=B\cdot v=0$, which yields that only three components of $E$ and three
components of $B$ are independent quantities.

$E$ and $B$ from (\ref{itf}) can be written as CBGQs in the $\{\gamma _{\mu
}\}$ basis and they are
\begin{eqnarray}
E &=&E^{\mu }\gamma _{\mu }=(1/c)F^{\mu \nu }v_{\nu }\gamma _{\mu }  \notag
\\
B &=&B^{\mu }\gamma _{\mu }=-(1/2c^{2})\varepsilon ^{\alpha \beta \nu \mu
}F_{\alpha \beta }v_{\nu }\gamma _{\mu }  \label{ab}
\end{eqnarray}
where $\varepsilon ^{\alpha \beta \nu \mu }$ is the totally skew-symmetric
Levi-Civita pseudotensor, $\varepsilon ^{0123}=1$.

When some reference frame is chosen and the standard basis $\{\gamma _{\mu
}\}$ in it and when $v$ is specified to be in the time direction in that
frame, i.e., $v=c\gamma _{0}$ (the $\gamma _{0}$ - system), which means that
the observers who measure the fields are at rest in that frame, then results
of the classical electromagnetism are recovered in that $\gamma _{0}$ -
system. Notice that we can select a particular, but otherwise arbitrary,
inertial frame of reference as the $\gamma _{0}$ - system, to which we shall
refer as the frame of our ``fiducial'' observers (for this name see [21]).
In the $\gamma _{0}$ - system equation (\ref{itf}) becomes
\begin{align}
F& =E_{f}\wedge \gamma _{0}+(\gamma _{5}B_{f})\cdot \gamma _{0}  \notag \\
E_{f}& =F\cdot \gamma _{0},\ B_{f}=-(1/c)\gamma _{5}(F\wedge \gamma _{0})
\label{ebg}
\end{align}
where in the $\{\gamma _{\mu }\}$ basis the pseudoscalar $I$ from (\ref{itf}%
) is $\gamma _{5}$, $I=\gamma _{5}$. The subscript $``f"$ in the above
relations (\ref{ebg}) stands for ``fiducial'' and denotes the explicit
dependence of these quantities on the $\gamma _{0}$ - observer, i.e.,
``fiducial'' - observer. It can be seen that in the $\gamma _{0}$ - system $%
E_{f}$ \emph{and} $B_{f}$ \emph{do not have the temporal components} $%
E_{f}^{0}=B_{f}^{0}=0$. Namely in the $\gamma _{0}$ - system with the $%
\{\gamma _{\mu }\}$ basis $E_{f}$ and $B_{f}$, written as CBGQs, are
\begin{eqnarray}
E_{f} &=&E_{f}^{\mu }\gamma _{\mu }=0\gamma _{0}+F^{i0}\gamma _{i}  \notag \\
B_{f} &=&B_{f}^{\mu }\gamma _{\mu }=0\gamma _{0}+(-1/2c)\varepsilon
^{0kli}F_{kl}\gamma _{i}.  \label{gl}
\end{eqnarray}
Thus $E_{f}$ and $B_{f}$ actually refer to the 3D subspace orthogonal to the
specific timelike direction $\gamma _{0}$. It is seen from (\ref{gl}) that
the components of $E_{f}$ and $B_{f}$ in the $\left\{ \gamma _{\mu }\right\}
$ basis are
\begin{equation}
E_{f}^{i}=F^{i0},\quad B_{f}^{i}=(-1/2c)\varepsilon ^{0kli}F_{kl}.
\label{sko}
\end{equation}
The relation (\ref{sko}) is nothing else than the standard identification of
the components $F^{\mu \nu }$ with the components of the 3D vectors $\mathbf{%
E}$ and $\mathbf{B}$, see, e.g., [2] equation (11.137) and the relations (%
\ref{ebg}) and (\ref{gl}) are, in fact, the spacetime split as in [7-10].

In Hestenes' decomposition of $F$, e.g., [9] equations (58)-(60), the
bivector field $F$ is expressed in terms of the sum of a relative vector $%
\mathbf{E}_{H}$ and a relative bivector $\gamma _{5}\mathbf{B}_{H}$ by
making a spacetime split in the $\gamma _{0}$ - system
\begin{align}
F& =\mathbf{E}_{H}+c\gamma _{5}\mathbf{B}_{H}\mathbf{,\quad E}_{H}=(F\cdot
\gamma _{0})\gamma _{0}  \notag \\
\mathbf{B}_{H}& =-(1/c)\gamma _{5}(F\wedge \gamma _{0})\gamma _{0}.
\label{FB}
\end{align}
where the subscript $H$ is for ``Hestenes.'' Both $\mathbf{E}_{H}$ and $%
\mathbf{B}_{H}$ are, in fact, bivectors. These relations, in the same way as
the relations (\ref{ebg}), \emph{are not manifestly Lorentz invariant}
equations; they are \emph{observer dependent }relations. The explicit
appearance of $\gamma _{0}$ in these expressions implies that \emph{the
spacetime split is observer dependent} and thus all quantities obtained by
the spacetime split in the $\gamma _{0}$ - system are \emph{observer
dependent quantities}. The difference between our approach and Hestenes' one
in electromagnetism is that Hestenes deals from the outset with the
spacetime split and the decomposition (\ref{FB}), while we start with
Lorentz invariant decomposition (\ref{itf}) and introduce the spacetime
split specifying the general velocity $v$ to be equal $c\gamma _{0}$.

This suggests that the relations (\ref{FB}) can also be made \emph{%
manifestly Lorentz invariant} equations, as are the equations (\ref{itf}),
by replacing $c\gamma _{0}$, the velocity of observers at rest, with some
general velocity $v$. Then the obtained equations are
\begin{align}
F& =E_{Hv}+cIB_{Hv}\mathbf{\quad }E_{Hv}=(1/c^{2})(F\cdot v)\wedge v  \notag
\\
B_{Hv}& =-(1/c^{3})I[(F\wedge v)\cdot v].  \label{he}
\end{align}
(The subscript $Hv$ is for ``Hestenes'' with $v$ and not, as usual [7-10],
with $\gamma _{0}$.) Now the relations (\ref{he}) completely correspond to
the equations (\ref{itf}). The relations (\ref{he}) were first presented in
[17, 18]. However, it is worth noting that it is much simpler and, in fact,
closer to the classical formulation of electromagnetism with the 3D $\mathbf{%
E}$ and $\mathbf{B}$ to work with the decomposition of $F$ into 1-vectors $E$
and $B$, as in (\ref{itf}), or in the $\gamma _{0}$ - system in (\ref{ebg}),
instead of decomposing $F$ into bivectors $E_{Hv}$ and $B_{Hv}$ (\ref{he}),
or in the $\gamma _{0}$ - system in (\ref{FB}). Thence we proceed using only
the decomposition of $F$ into 1-vectors $E$ and $B$ (\ref{itf}), or (\ref
{ebg}).\bigskip \medskip

\noindent \textbf{5. }$K_{L}$, $M$, $N$ \textbf{as\ 4D AQs\ or\ 4D
CBGQs\bigskip }

\noindent All quantities that appear in the problem discussed by Jackson [1]
can be written as 4D AQs and equations with them will be \emph{manifestly
Lorentz invariant} equations. Thus the position 1-vector in the 4D spacetime
is $x$. Then $x=x(\tau )$ determines the history of a particle with proper
time $\tau $ and proper velocity $u=dx/d\tau $. The Lorentz force as a 4D AQ
(1-vector) is $K_{L}=(q/c)F\cdot u$, where $u$ is the velocity 1-vector of a
charge $q$ (it is defined to be the tangent to its world line). In the usual
geometric algebra approaches [7-10] to SR one makes from the outset the
spacetime split and writes the Lorentz force $K_{L}$ (1-vector) in the Pauli
algebra of $\gamma _{0}$. Since this procedure is observer dependent we
express $K_{L}$ in terms of AQs 1-vectors $E$ and $B$ as
\begin{equation}
K_{L}=(q/c)F\cdot u=(q/c)\left[ (1/c)E\wedge v+(IB)\cdot v\right] \cdot u
\label{KEB}
\end{equation}
see also [15,17]. (Of course the whole consideration could be equivalently
made using $E_{Hv}$ and $B_{Hv}$ from (\ref{he}) but with more complicated
expressions.) The equivalent expression in the tensor formalism, \emph{with
tensors as AQs}, is given, e.g., in [20], by Vanzella, Matsas and Crater. In
the general case when charge and observer have distinct worldlines the
Lorentz force $K_{L}$ (\ref{KEB}) can be written as a sum of the $v-\perp $
part $K_{L\perp }$ and the $v-\parallel $ part $K_{L\parallel },$ $%
K_{L}=K_{L\perp }+K_{L\parallel },$ where
\begin{equation}
K_{L\perp }=(q/c^{2})(v\cdot u)E+(q/c)((IB)\cdot v)\cdot u  \label{Kaok}
\end{equation}
\begin{equation}
K_{L\parallel }=(-q/c^{2})(E\cdot u)v  \label{kapa}
\end{equation}
respectively. Of course $K_{L}$, $K_{L\perp }$ and $K_{L\parallel }$ are all
4D quantities defined without reference frames, the AQs, and the
decomposition of $K_{L}$ into $K_{L\perp }$ and $K_{L\parallel }$ is an
observer independent decomposition. It can be easily verified that $%
K_{L\perp }\cdot v=0$ and $K_{L\parallel }\wedge v=0$. Particularly from the
definition of the Lorentz force $K_{L}=(q/c)F\cdot u$ and the relation $%
E=(1/c)F\cdot v$ (from (\ref{itf})) it follows that the Lorentz force
ascribed by an observer comoving with a charge, $u=v$, is \emph{purely
electric }$K_{L}=qE$.

Both parts of $K_{L}$ can be written as CBGQs in the standard basis $%
\{\gamma _{\mu }\}$
\begin{equation}
K_{L\perp }=(q/c^{2})(v^{\nu }u_{\nu })E^{\mu }\gamma _{\mu }+(q/c)%
\widetilde{\varepsilon }_{\ \nu \rho }^{\mu }u^{\nu }B^{\rho }\gamma _{\mu }
\label{kc}
\end{equation}
where $\widetilde{\varepsilon }_{\mu \nu \rho }\equiv \varepsilon _{\lambda
\mu \nu \rho }v^{\lambda }$ is the totally skew-symmetric Levi-Civita
pseudotensor induced on the hypersurface orthogonal to $v$ and
\begin{equation}
K_{L\parallel }=(-q/c^{2})(E^{\nu }u_{\nu })v^{\mu }\gamma _{\mu }.
\label{ki}
\end{equation}
Speaking in terms of the prerelativistic notions one can say that in the
approach with the 1-vectors $E$ and $B$ $K_{\perp }$ plays the role of the
usual Lorentz force lying on the 3D hypersurface orthogonal to $v$, while $%
K_{\parallel }$ is related to the work done by the field on the charge.
However \emph{in our invariant SR only both components together, equations }(%
\ref{Kaok})\emph{\ and }(\ref{kapa})\emph{, have physical meaning and they
define the Lorentz force both in the theory and in experiments. }

Further the angular momentum $M$ (bivector), the torque $N$ (bivector) about
the origin for some force $K$ (1-vector) and manifestly Lorentz invariant
equation connecting $M$ and $N$ are defined as
\begin{eqnarray}
M &=&x\wedge p,\ p=mu,  \notag \\
N &=&x\wedge K;\quad N=dM/d\tau  \label{MKN}
\end{eqnarray}
where for the Lorentz force $K_{L}$ the torque $N$ about the origin becomes $%
N=x\wedge K_{L}$.

When $M$ and $N$ (for the Lorentz force $K_{L}$) are written as CBGQs in the
$\{\gamma _{\mu }\}$ basis they become
\begin{eqnarray}
M &=&(1/2)M^{\mu \nu }\gamma _{\mu }\wedge \gamma _{\nu },\ M^{\mu \nu
}=m(x^{\mu }u^{\nu }-x^{\nu }u^{\mu }),  \notag \\
N &=&(1/2)N^{\mu \nu }\gamma _{\mu }\wedge \gamma _{\nu },\ N^{\mu \nu
}=x^{\mu }K_{L}^{\nu }-x^{\nu }K_{L}^{\mu }.  \label{mn}
\end{eqnarray}
We see that the components $M^{\mu \nu }$ ($M^{\alpha \beta }=\gamma ^{\beta
}\cdot (\gamma ^{\alpha }\cdot M)$) from (\ref{mn}) are identical to the
covariant angular momentum four-tensor given by equation (A3) in Jackson's
paper [1]. However $M$ and $N$ from (\ref{MKN}) are geometric 4D quantities,
the AQs, which are \emph{independent of the chosen reference frame and of
the chosen system of coordinates in it}, whereas the components $M^{\mu \nu
} $ and $N^{\mu \nu }$ that are used in the usual covariant approach, e.g.,
equation (A3) in [1], are coordinate quantities, the numbers obtained in the
specific system of coordinates, Einstein's system of coordinates, i.e., in
the $\{\gamma _{\mu }\}$ basis. Notice that, in contrast to the usual
covariant approach, $M$ and $N$ from (\ref{mn}) are also geometric 4D
quantities, the CBGQs, which contain both components \emph{and a basis},
here bivector basis $\gamma _{\mu }\wedge \gamma _{\nu }$.

It is worth noting that the principle of relativity is automatically
included in such a theory with geometric 4D quantities, AQs or CBGQs,
whereas in the standard approach to SR [4] the principle of relativity is
postulated outside the framework of a mathematical formulation of the
theory. \bigskip \medskip

\noindent \textbf{6. The\ LT of 4D\ }$E$\textbf{\ and\ }$B$\textbf{\ and of
other multivectors}\bigskip

\noindent In the usual Clifford algebra formalism [7-11] the LT are
considered as active transformations acting on multivectors as AQs. When AQs
are written as CBGQs in some basis then the components of, e.g., some
1-vector relative to a given inertial frame of reference (with the standard
basis $\left\{ \gamma _{\mu }\right\} $) are transformed by the active LT
into the components of a new 1-vector relative to the same frame (the basis $%
\left\{ \gamma _{\mu }\right\} $ is not changed). Furthermore the LT are
described with rotors $R,$ $R\widetilde{R}=1$, in the usual way as $%
p\rightarrow p^{\prime }=Rp\widetilde{R}$ and it is $=p_{\mu }^{\prime
}\gamma ^{\mu }$ when the $\left\{ \gamma _{\mu }\right\} $ basis is
introduced. To an observer in the $\left\{ \gamma _{\mu }\right\} $ basis
the vector $p^{\prime }$ appears the same as the vector $p$ appears to an
observer in the $\left\{ \gamma _{\mu }^{\prime }\right\} $ basis.
(Reversion is an invariant kind of conjugation, which is defined by $%
\widetilde{AB}=\widetilde{B}\widetilde{A},$ $\widetilde{a}=a$ for any vector
$a$, and it reverses the order of vectors in any given expression.) But
every rotor in spacetime can be written in terms of a bivector as $%
R=e^{\theta /2}.$ For boosts in arbitrary direction
\begin{equation}
R=e^{\theta /2}=(1+\gamma -\gamma \beta \gamma _{0}n)/(2(1+\gamma ))^{1/2},
\label{err}
\end{equation}
$\theta =\alpha \gamma _{0}n,$ $\beta $ is the scalar velocity in units of $%
c $, $\gamma =(1-\beta ^{2})^{-1/2}$, or in terms of an `angle' $\alpha $ we
have $\tanh \alpha =\beta ,$ $\cosh \alpha =\gamma ,$ $\sinh \alpha =\beta
\gamma ,$ and $n$ is not the basis vector but any unit space-like vector
orthogonal to $\gamma _{0};$ $e^{\theta }=\cosh \alpha +\gamma _{0}n\sinh
\alpha .$ One can also express the relationship between the two relatively
moving frames $S$ and $S^{\prime }$ in terms of rotor as $\gamma _{\mu
}^{\prime }=R\gamma _{\mu }\widetilde{R}.$ For boosts in the direction $%
\gamma _{1}$ the rotor $R$ is given by the relation (\ref{err}) with $\gamma
_{1}$ replacing $n$ (all in the standard basis $\left\{ \gamma _{\mu
}\right\} $). Then \emph{for any multivector} $M$ the active LT are defined
by the relation
\begin{equation}
M^{\prime }=RM\widetilde{R}.  \label{aLT}
\end{equation}

When the active LT (\ref{aLT}) are applied to 1-vectors $E_{f}$ and $B_{f}$
from Eq. (\ref{gl}) one finds the transformed $E_{f}^{\prime }$ as
\begin{align}
E_{f}^{\prime }& =R(F\cdot \gamma _{0})\widetilde{R}=(RF\widetilde{R})\cdot
(R\gamma _{0}\widetilde{R})=F^{\prime }\cdot \gamma _{0}^{\prime }=  \notag
\\
R(F^{k0}\gamma _{k})\widetilde{R}& =E_{f}^{\prime \mu }\gamma _{\mu }=-\beta
\gamma E_{f}^{1}\gamma _{0}+\gamma E_{f}^{1}\gamma _{1}+E_{f}^{2}\gamma
_{2}+E_{f}^{3}\gamma _{3},  \label{nle}
\end{align}
which is the usual form for the active LT of the 1-vector $E_{f}=E_{f}^{\mu
}\gamma _{\mu }$. Similarly we find for $B_{f}^{\prime }$
\begin{align}
B_{f}^{\prime }& =R\left[ -(1/c^{2})\gamma _{5}(F\wedge c\gamma _{0})\right]
\widetilde{R}=R\left[ (-1/2c)\varepsilon ^{0kli}F_{kl}\gamma _{i}\right]
\widetilde{R}=  \notag \\
& =B_{f}^{\prime \mu }\gamma _{\mu }=-\beta \gamma B_{f}^{1}\gamma
_{0}+\gamma B_{f}^{1}\gamma _{1}+B_{f}^{2}\gamma _{2}+B_{f}^{3}\gamma _{3},
\label{nlb}
\end{align}
which is the familiar form for the active LT of the 1-vector $%
B_{f}=B_{f}^{\mu }\gamma _{\mu }$. It is important to note

(i) \emph{that} $E_{f}^{\prime }$ \emph{and} $B_{f}^{\prime }$ \emph{are not
orthogonal to} $\gamma _{0},$ i.e., \emph{they} \emph{have} \emph{temporal
components} $\neq 0.$ \emph{They do not belong to the same 3D subspace as} $%
E_{f}$ \emph{and} $B_{f},$ \emph{but they are in the 4D spacetime spanned by
the whole standard basis }$\left\{ \gamma _{\mu }\right\} $.

The relations (\ref{nle}) and (\ref{nlb}) imply that the spacetime split in
the $\gamma _{0}$ - system is not possible for the transformed $F^{\prime
}=RF\widetilde{R}$, i.e., $F^{\prime }$ cannot be decomposed into $%
E_{f}^{\prime }$ and $B_{f}^{\prime }$ as $F$ is decomposed in the relation (%
\ref{ebg}), $F^{\prime }\neq E_{f}^{\prime }\wedge \gamma _{0}+c(\gamma
_{5}B_{f}^{\prime })\cdot \gamma _{0}.$ Notice, what is very important, that

(ii) \emph{the components} $E_{f}^{\mu }$ ($B_{f}^{\mu }$) \emph{from
equation} (\ref{gl}) \emph{transform upon the active LT again to the
components} $E_{f}^{\prime \mu }$ ($B_{f}^{\prime \mu }$) \emph{from
equations }(\ref{nle}) ((\ref{nlb})); \emph{there is no mixing of components}%
. \emph{Thus} \emph{by the active LT} $E_{f}$ \emph{transforms to} $%
E_{f}^{\prime }$ \emph{and} $B_{f}$ \emph{to }$B_{f}^{\prime }.$

Actually, as we said, this is the way in which every 1-vector transforms
upon the active LT. The LT of the 4D $E$ and $B$ in the tensor formalism are
already presented in [16] and in geometric algebra formalism in [17, 18].

The same results can be obtained with the passive LT, either by using a
coordinate-free form of the LT (such one as in [12,13,15]), or by using the
standard expressions for the matrix of the LT in the Einstein system of
coordinates from, e.g., [2], see also the discussion about passive and
active LT in Hestenes' paper [9] and equations (93) - (95) therein. \emph{%
The passive LT always transform the whole 4D quantity, basis and components,
leaving the whole 4D quantity unchanged.} Thus under the passive LT the
field bivector $F$ as a well-defined 4D quantity remains unchanged, i.e., $%
F=(1/2)F^{\mu \nu }\gamma _{\mu }\wedge \gamma _{\nu }=(1/2)F^{\prime \mu
\nu }\gamma _{\mu }^{\prime }\wedge \gamma _{\nu }^{\prime }$ (all primed
quantities are the Lorentz transforms of the unprimed ones). In the same way
it holds that, e.g., $E_{f}^{\mu }\gamma _{\mu }=E_{f}^{\prime \mu }\gamma
_{\mu }^{\prime }$. The invariance of some 4D CBGQ upon the passive LT
reflects the fact that such mathematical, invariant, geometric 4D quantity
represents \emph{the same physical object} for relatively moving observers.
Thus in the invariant SR we consider that \emph{quantity which does not
change upon the passive LT has an independent physical reality, both
theoretically and experimentally. }

The importance of the \emph{concept of sameness} of a physical system for
different observers is first emphasized in papers by Rohrlich [25] and Gamba
[26] and further developed and clarified in, e.g., [12-14], where it is
proved that the Lorentz contraction and the dilatation of time belong to the
``apparent'' transformations and not to the ``true'' transformations. The
``apparent'' transformations do not refer to the same quantity in the 4D
spacetime but to the same measurements, whereas the ``true''
transformations, as are the LT, refer to the same 4D quantity. For example,
as explained in [12-14], in the Lorentz contraction the rest spatial length $%
L_{0}$ of a rod in its rest frame $S$ and the spatial length $L^{\prime }$
of that rod in relatively moving inertial frame $S^{\prime }$ do not refer
to the same 4D tensor quantity but to two different quantities in 4D
spacetime. These quantities are obtained by the same measurements in $S$ and
$S^{\prime }$; the spatial ends of the rod are measured simultaneously at
some $t=a$ in $S$ and also at some $t^{\prime }=b$ in $S^{\prime }$, and $a$
in $S$ and $b$ in $S^{\prime }$ are not related by the LT or any other
coordinate transformation; see figure 3 in [13] and compare it with figure 1
in [13] for the correct 4D geometric quantity, the spacetime length for a
moving rod. The names ``apparent'' and ``true'' transformations are
introduced in Rohrlich's paper [25]. The comparisons [14] with well-known
experiments that test SR as are the Michelson-Morley experiment, the
''muon'' experiments, the Kennedy-Thorndike type experiments and the
Ives-Stilwell type experiments explicitly show that all these experiments
are in a complete agreement with the geometric approach of the invariant SR,
whereas, contrary to the general belief, it is not the case for the usual
approach that deals with the ``apparent'' transformations, the Lorentz
contraction and the dilatation of time.\bigskip \medskip

\noindent \textbf{7. The derivation of the ST of the electric and magnetic
fields\bigskip }

\noindent Let us now see how the ST (\ref{Aeb}) or (\ref{A1}) are obtained
in a rigorous mathematical way from the geometric approach to SR, .i.e., in
the invariant SR. \emph{The ST for} $E_{st}^{\prime }$ \emph{and} $%
B_{st}^{\prime }$ (the subscript $st$ is for standard)\emph{\ are derived
assuming that the quantities obtained by the active LT of} $E_{f}$ \emph{and}
$B_{f}$ \emph{are again in the 3D subspace of the} $\gamma _{0}$ \emph{-}
\emph{observer}. Thus

(i') \emph{it is supposed that for the transformed} $E_{st}^{\prime }$ \emph{%
and} $B_{st}^{\prime }$ \emph{again hold that} $E_{st}^{\prime
0}=B_{st}^{\prime 0}=0,$ \emph{i.e., that} $E_{st}^{\prime }\cdot \gamma
_{0}=B_{st}^{\prime }\cdot \gamma _{0}=0$ \emph{as for} $E_{f}$ \emph{and }$%
B_{f}.$

Thence, in contrast to the LT of $E_{f}$ and $B_{f},$ (\ref{nle}) and (\ref
{nlb}) respectively, it is assumed in all Clifford algebra formalisms, e.g.,
[7-11], that
\begin{align}
E_{st}^{\prime }& =(RF\widetilde{R})\cdot \gamma _{0}=F^{\prime }\cdot
\gamma _{0}=F^{\prime i0}\gamma _{i}=E_{st}^{\prime i}\gamma _{i}=  \notag \\
& =E_{f}^{1}\gamma _{1}+\gamma (E_{f}^{2}-\beta cB_{f}^{3})\gamma
_{2}+\gamma (E_{f}^{3}+\beta cB_{f}^{2})\gamma _{3},  \label{ce}
\end{align}
where $F^{\prime }=RF\widetilde{R}$. Similarly we find for $B_{st}^{\prime }$
\begin{align}
B_{st}^{\prime }& =-(1/c)\gamma _{5}(F^{\prime }\wedge \gamma
_{0})=-(1/2c)\varepsilon ^{0kli}F_{kl}^{\prime }\gamma _{i}=B_{st}^{\prime
i}\gamma _{i}=  \notag \\
& B_{f}^{1}\gamma _{1}+(\gamma B_{f}^{2}+\beta \gamma E_{f}^{3}/c)\gamma
_{2}+(\gamma B_{f}^{3}-\beta \gamma E_{f}^{2}/c)\gamma _{3}.  \label{B}
\end{align}
The ST of, e.g., $E_{f}$, are given by (\ref{ce}) and this relation shows
\emph{that only} $F$ \emph{is transformed while }$\gamma _{0}$ \emph{is not
transformed. }This is the fundamental difference between the LT (\ref{nle})
and (\ref{nlb}) and the ST (\ref{ce}) and (\ref{B}). From the
transformations (\ref{ce}) and (\ref{B}) one simply finds the
transformations of the spatial components $E_{st}^{\prime i}$ and $%
B_{st}^{\prime i}$
\begin{equation}
E_{st}^{\prime i}=F^{\prime i0},\quad B_{st}^{\prime i}=(-1/2c)\varepsilon
^{0kli}F_{kl}^{\prime },  \label{sk1}
\end{equation}
which is the relation (\ref{sko}) with the primed quantities. As can be seen
from equations (\ref{ce}), (\ref{B}) and (\ref{sk1}) \emph{the
transformations for} $E_{st.}^{\prime i}$ \emph{and} $B_{st.}^{\prime i}$%
\emph{\ are} \emph{the ST of components of the 3D vectors} $\mathbf{E}$
\emph{and} $\mathbf{B}$, equation (\ref{A1}) (and for the 3D $\mathbf{E}$
and $\mathbf{B}$, equation (\ref{Aeb})), which are quoted in almost every
textbook and paper on relativistic electrodynamics including [3-5], see,
e.g. Jackson's book [2] section 11.10. These relations (\ref{ce}), (\ref{B})
and (\ref{sk1}) are explicitly derived and given in the Clifford algebra
formalism, e.g., in [7] equation (18.22), [8] chapter 9 equations (3.51a,b),
[10] equation (7.33) and in [11] chapter 7 equations (20a,b). Notice that,
in contrast to the active LT (\ref{nle}) and (\ref{nlb}),

(ii') \emph{according to the ST} (\ref{ce}) \emph{and }(\ref{B}) (\emph{i.e.}%
, (\ref{sk1})) \emph{the transformed components} $E_{st}^{\prime i}$ \emph{%
are expressed by the mixture of components} $E_{f}^{i}$ \emph{and} $%
B_{f}^{i},$ \emph{and the same holds for} $B_{st}^{\prime i}$.

In all previous treatments of SR the transformations for $E_{st.}^{\prime i}$
and $B_{st.}^{\prime i}$ are considered to be the LT of the 3D electric and
magnetic fields. However our analysis shows that the transformations for $%
E_{st.}^{\prime i}$ and $B_{st.}^{\prime i}$, equation (\ref{sk1}), are
derived from the transformations (\ref{ce}) and (\ref{B}), which differ from
the LT; the LT are given by the relations (\ref{nle}) and (\ref{nlb}).

What is with the \emph{concept of sameness} when the ST (\ref{ce}) and (\ref
{B}), i.e., (\ref{Aeb}) or (\ref{A1}), are used. It can be easily shown that
$E_{f}^{\mu }\gamma _{\mu }\neq E_{st}^{\prime \mu }\gamma _{\mu }^{\prime
}. $ This means that, e.g., $E_{f}^{\mu }\gamma _{\mu }$ and $%
E_{st.}^{\prime \mu }\gamma _{\mu }^{\prime }$ \emph{are not the same
quantity for observers in} $S$ \emph{and} $S^{\prime },$ and that \emph{the
ST are also the ``apparent'' transformations}. As far as relativity is
concerned the quantities, e.g., $E_{f}^{\mu }\gamma _{\mu }$ and $%
E_{st.}^{\prime \mu }\gamma _{\mu }^{\prime },$ are not related to one
another. The fact that they are measured by two observers ($\gamma _{0}$ -
and $\gamma _{0}^{\prime }$ - observers) does not mean that relativity has
something to do with the problem. The reason is that observers in the $%
\gamma _{0}$ - system and in the $\gamma _{0}^{\prime }$ - system are not
looking at the same 4D physical object but at two different 4D objects.
\emph{Every observer makes measurement on its own object and such
measurements are not related by the LT.} Thus the transformations for $%
E_{st.}^{\prime i}$ and $B_{st.}^{\prime i}$, (\ref{ce}), (\ref{B}) and (\ref
{sk1}) or (\ref{Aeb}) and (\ref{A1}), are not the same as the LT of
well-defined 4D quantities, (\ref{nle}) and (\ref{nlb}). (All these results
are presented in the tensor formalism in [16] and in the geometric algebra
formalism in [17-19], where they are also compared with experiments.)

The knowledge of this fundamental difference between the ST and the LT
enables us to resolve in a simple way Jackson's paradox [1] that there is a
torque and so a time rate of change of angular momentum in one inertial
frame, but no angular momentum and no torque in another. \bigskip \medskip

\noindent \textbf{8. The resolution\ of\ the\ paradox\bigskip }

\noindent First let us formulate the problem using AQs. The torque $N$ about
the origin as an AQ is $N=x\wedge K_{L}$, where $K_{L}$ is the Lorentz force
given by (\ref{KEB}) or (\ref{Kaok})\ and (\ref{kapa}). $E$ and $B$ for a
charge $Q$ moving with constant velocity $u_{Q}$ can be determined from (\ref
{itf}) and the expression for the electromagnetic field $F$ (\ref{cvf}).
They are
\begin{eqnarray}
E &=&(D/c^{2})[(u_{Q}\cdot v)x-(x\cdot v)u_{Q}]  \notag \\
B &=&(-D/c^{3})I(x\wedge u_{Q}\wedge v),  \label{ec}
\end{eqnarray}
where $D=kQ/\left| x\wedge (u_{Q}/c)\right| ^{3}$ and, as before, $k=1/4\pi
\varepsilon _{0}$. (The relation (\ref{ec}) is already derived in [19].) All
these quantities are AQs, i.e., they are independent of the chosen reference
frame and of the chosen system of coordinates in it. When the world lines of
the observer and the charge $Q$ coincide, $u_{Q}=v$, then (\ref{ec}) yields
that $B=0$ and only an electric field (Coulomb field) remains.

The next step is to write all AQs as CBGQs in some conveniently chosen
inertial frame with an appropriate basis in it. \emph{The main advantage of
such geometric approach is that when CBGQs are determined in a chosen
inertial frame they remain unchanged in all other relatively moving inertial
frames and they are independent of the chosen system of coordinates in these
frames. }

In our case one choice for the starting, convenient, frame is the $S^{\prime
}$ frame, in which a point charge $Q$ is fixed permanently at the origin ($%
u_{Q}=c\gamma _{0}^{\prime }$), and in that frame let the observers who
measure the fields are at rest, i.e., in $S^{\prime }$, $v=c\gamma
_{0}^{\prime }$ in (\ref{ec}). Thus the $S^{\prime }$ frame is the frame of
our ``fiducial'' observers or the $\gamma _{0}$ - system in which results of
the classical electromagnetism with the 3D $\mathbf{E}$ and $\mathbf{B}$ are
recovered. However in contrast to the classical electromagnetism we are not
concerned with the 3D $\mathbf{E}$ and $\mathbf{B}$ than by the 4D $E$ and $%
B $, which have only spatial components in the frame of ``fiducial''
observers. (Notice that, as already said, the results do not depend on our
choice for the $\gamma _{0}$ - system.) Further in $S^{\prime }$ we choose
Einstein's system of coordinates, that is, the $\{\gamma _{\mu }\}$ basis.
When we show that the torque $N=(1/2)N^{^{\prime }\mu \nu }\gamma _{\mu
}^{\prime }\wedge \gamma _{\nu }^{\prime }=0$ in $S^{\prime }$ then due to
the invariance of any CBGQ upon the passive LT $N$ \emph{will be zero} \emph{%
in all other relatively moving inertial frames, }thus in the laboratory
frame, the $S$ frame, as well,
\begin{equation}
N=(1/2)N^{^{\prime }\mu \nu }\gamma _{\mu }^{\prime }\wedge \gamma _{\nu
}^{\prime }=(1/2)N^{\mu \nu }\gamma _{\mu }\wedge \gamma _{\nu }=0.
\label{enc}
\end{equation}
The paradox does not appear since the principle of relativity is
automatically satisfied in such an approach to SR which exclusively deals
with geometric 4D quantities, i.e., AQs or CBGQs.\bigskip \medskip

\noindent \textit{8.1. The proof that all}\textbf{\ }$N^{^{\prime }\mu \nu
}=0$\textit{\ in}\textbf{\ }$S^{\prime }\bigskip $

\noindent Now let us show that all components $N^{^{\prime }\mu \nu }$ are
zero in $S^{\prime }$. $N^{\prime \mu \nu }=x^{\prime \mu }K_{L}^{\prime \nu
}-x^{\prime \nu }K_{L}^{\prime \mu }$. The components $x^{\prime \mu }$ are
as in section 2 but we write them without the subscript ``$A$,'' $x^{\prime
\mu }=(ct^{\prime }=0,x^{\prime 1},x^{\prime 2},0)$. The components of the
Lorentz force $K_{L}$ are determined from the relations (\ref{kc}) for $%
K_{L\perp }$ and (\ref{ki}) for $K_{L\parallel }$. The electric and magnetic
fields, $E$ and $B$ respectively, are determined from the relation (\ref{ec}%
) taking into account that in $S^{\prime }$ $v=c\gamma _{0}^{\prime }$,
which yields that their temporal components are zero in $S^{\prime }$ (as in
(\ref{gl})). (The $S^{\prime }$ frame is the frame of ``fiducial''
observers.)\textbf{\ }Further in $S^{\prime }$ $u_{Q}=c\gamma _{0}^{\prime }$
as well, which, from (\ref{ec}), yields, as already said, that the whole $%
B=0 $; $B=B^{^{\prime }\mu }\gamma _{\mu }^{\prime }=0$. (Notice that due to
invariance of any CBGQ upon the passive LT \emph{the magnetic field} $B^{\mu
}\gamma _{\mu }=0$ \emph{in the laboratory frame} $S$ \emph{too.}) The
electric field is
\begin{equation}
E=E^{\prime \mu }\gamma _{\mu }^{\prime }=D(x^{\prime 1}\gamma _{1}^{\prime
}+x^{\prime 2}\gamma _{2}^{\prime }),  \label{fc}
\end{equation}
where $D=kQ/r^{\prime 3}$. Of course, the spatial components of $E$ are the
same as the components of $\mathbf{E}^{\prime }(t_{A}^{\prime }=0)$ from
section 2 as it must. In $S^{\prime }$ the velocity 1-vector of the charge $%
q $ (at $t^{\prime }=0$) is $u=c\gamma _{0}^{\prime }$, i.e., $%
u=v(=u_{Q})=c\gamma _{0}^{\prime }$; in $S^{\prime }$ both charges $Q$ and $%
q $ are at rest. This yields that in $K_{L}$, which is purely electric, $%
K_{L\parallel }=0$ and $K_{L\perp }=q(E^{\prime 1}\gamma _{1}^{\prime
}+E^{\prime 2}\gamma _{2}^{\prime })$. Thus it holds that
\begin{equation}
K_{L}=K_{L\perp }=qE=qE^{\prime \mu }\gamma _{\mu }^{\prime }  \label{kli}
\end{equation}
and it is $=qE^{\mu }\gamma _{\mu }$ in $S$. Then the torque $N$ becomes
\begin{equation}
N=(x^{\prime 1}K_{L}^{\prime 2}-x^{\prime 2}K_{L}^{\prime 1})(\gamma
_{1}^{\prime }\wedge \gamma _{2}^{\prime })=qD(x^{\prime 1}x^{\prime
2}-x^{\prime 2}x^{\prime 1})(\gamma _{1}^{\prime }\wedge \gamma _{2}^{\prime
})=0.  \label{n0}
\end{equation}
Taking into account the relations (\ref{enc}) and (\ref{n0}) we conclude
that there is no violation of the principle of relativity and consequently
the paradox does not appear in our approach with geometric 4D
quantities.\bigskip \medskip

\noindent \textit{8.2. The proof that}\textbf{\ }$N^{\mu \nu }=0$ \textit{in}%
\textbf{\ }$S$ \textit{using the LT of}\textbf{\ }$K_{L}$ \textit{and} $x$ $%
\bigskip $

\noindent Although the relations (\ref{enc}) and (\ref{n0}) complete the
proof that the torque $N$ is zero in all relatively moving inertial frames
if it is zero in any one of them we shall, for readers' convenience,
explicitly show that the torque $N$ is zero in the laboratory frame, the $S$
frame, if it is zero in the $S^{\prime }$ frame. This can be shown in
different ways.

One way is to explicitly show that all $N^{\mu \nu }=0$ when $N^{^{\prime
}\mu \nu }=0$ using directly the passive LT of the CBGQs $K_{L}^{\mu }\gamma
_{\mu }$ and $x^{\mu }\gamma _{\mu }$. The components of $N$ in the $S$
frame are $N^{\mu \nu }=x^{\mu }K_{L}^{\nu }-x^{\nu }K_{L}^{\mu }$, where
the components $x^{\mu }$ and $K_{L}^{\mu }$
\begin{eqnarray}
x^{\mu } &=&(\gamma \beta x^{\prime 1},\gamma x^{\prime 1},x^{\prime 2},0),
\notag \\
K_{L}^{\mu } &=&(\gamma \beta K_{L}^{\prime 1},\gamma K_{L}^{\prime
1},K_{L}^{\prime 2},0),  \label{k1}
\end{eqnarray}
are obtained by the LT from $x^{\prime \mu }=(0,x^{\prime 1},x^{\prime 2},0)$
and $K_{L}^{\prime \mu }=(0,K_{L}^{\prime 1},K_{L}^{\prime 2},0)$. In fact,
the whole CBGQs $x^{\prime \mu }\gamma _{\mu }^{\prime }$ and $K_{L}^{\prime
\mu }\gamma _{\mu }^{\prime }$ are transformed by the passive LT from $%
S^{\prime }$ to $S$, and it holds that $x=x^{\prime \mu }\gamma _{\mu
}^{\prime }=x^{\mu }\gamma _{\mu }$ and similarly for $K_{L}$. Then it is
easy to see that all components $N^{\mu \nu }$ are zero except $%
N^{02}=(-N^{20})=\gamma \beta (x^{\prime 1}K_{L}^{\prime 2}-x^{\prime
2}K_{L}^{\prime 1})$ and $N^{12}=(-N^{21})=\gamma (x^{\prime 1}K_{L}^{\prime
2}-x^{\prime 2}K_{L}^{\prime 1})$, but due to (\ref{n0}) they are also zero,
whence it follows that all $N^{\mu \nu }=0$ and consequently $N=(1/2)N^{\mu
\nu }\gamma _{\mu }\wedge \gamma _{\nu }=0$.\bigskip \medskip

\noindent \textit{8.3. The proof that}\textbf{\ }$N^{\mu \nu }=0$ \textit{in}%
\textbf{\ }$S$ \textit{using the LT of}\textbf{\ }$E$ \textit{and }$B$.
\textit{The frame}

\textit{of ``fiducial'' observers is the}\textbf{\ }$S^{\prime }$ \textit{%
frame}$\bigskip $

\noindent Another way is, e.g., to use the passive LT corresponding to the
active ones (\ref{nle}) for the transformations of CBGQs $E^{\prime \mu
}\gamma _{\mu }^{\prime }$ and $B^{\prime \mu }\gamma _{\mu }^{\prime }$ to $%
E^{\mu }\gamma _{\mu }$ and $B^{\mu }\gamma _{\mu }$. We suppose, as above,
that the observers who measure the fields are at rest in $S^{\prime }$,
i.e., $v=c\gamma _{0}^{\prime }$, thus $v^{\prime \mu }=(c,0,0,0)$ in $%
S^{\prime }$. (It is already mentioned that with this choice $%
u=v(=u_{Q})=c\gamma _{0}^{\prime }$.) The CBGQ $E^{\prime \mu }\gamma _{\mu
}^{\prime }$ is given by (\ref{fc}) and all $B^{\prime \mu }$ are zero,
i.e., $B=B^{\prime \mu }\gamma _{\mu }^{\prime }=0$. The CBGQs $E^{\mu
}\gamma _{\mu }$ and $B^{\mu }\gamma _{\mu }$ in $S$ are determined by the
passive LT of fields (corresponding to the active LT (\ref{nle})), whence
the components in $S$ are
\begin{equation}
E^{\mu }=(\gamma \beta E^{\prime 1},\gamma E^{\prime 1},E^{\prime
2},0),\quad B^{\mu }=0.  \label{f0}
\end{equation}
Notice that in $S$ there is a \emph{temporal component} $E^{0}=\gamma \beta
E^{\prime 1}$ and \emph{there is no magnetic field} in relatively moving
inertial frame $S$ if it was zero in the frame of ``fiducial'' observers,
here the\textbf{\ }$S^{\prime }$ frame. This is, as already mentioned, a
fundamental difference relative to the ST (\ref{ce}) and (\ref{B}), i.e., (%
\ref{Aeb}) or (\ref{A1}). Remember that upon the passive LT the unit
1-vectors $\gamma _{\mu }^{\prime }$ transform to $\gamma _{\mu }$ and it
holds that $E=E^{\prime \mu }\gamma _{\mu }^{\prime }=E^{\mu }\gamma _{\mu }$
(the same quantity for observers in $S^{\prime }$ and $S$) and also $%
B=B^{\prime \mu }\gamma _{\mu }^{\prime }=B^{\mu }\gamma _{\mu }=0$. When $%
K_{L}$ is written as a CBGQ in $S$ and in the $\{\gamma _{\mu }\}$ basis it
is given as the sum of (\ref{kc}) and (\ref{ki})
\begin{equation}
K_{L}=(q/c^{2})[(v^{\nu }u_{\nu })E^{\mu }+\widetilde{\varepsilon }_{\ \nu
\rho }^{\mu }u^{\nu }cB^{\rho }-(E^{\nu }u_{\nu })v^{\mu }]\gamma _{\mu }.
\label{lo}
\end{equation}
Now comes an important point. The CBGQs $v^{\mu }\gamma _{\mu }$ and $u^{\mu
}\gamma _{\mu }$ in $S$ are also determined by the passive LT from those in $%
S^{\prime }$; the observers who were at rest in $S^{\prime }$, the
``fiducial'' observers, are now moving in $S$, and the charge $q$ is also
moving in $S$. Thence in $S$ the components are
\begin{equation}
v^{\mu }(=u^{\mu })=(\gamma c,\gamma \beta c,0,0)  \label{ev}
\end{equation}
(for the whole CBGQ it again holds $v=v^{\prime \mu }\gamma _{\mu }^{\prime
}=v^{\mu }\gamma _{\mu }$ and the same for $u$). Equation (\ref{ev})
together with (\ref{lo}) leads to
\begin{eqnarray}
K_{L}^{0} &=&q(1-\gamma ^{2})E^{0}+q\beta \gamma ^{2}E^{1},  \notag \\
K_{L}^{1} &=&q(1+\beta ^{2}\gamma ^{2})E^{1}-q\beta \gamma ^{2}E^{0},
\label{ok} \\
K_{L}^{2} &=&qE^{2},\quad K_{L}^{3}=qE^{3}.  \notag
\end{eqnarray}
It is worth noting that the magnetic field $B$ does not appear in the
Lorentz force. Such result for $B$ is obtained not only in $S^{\prime }$,
the frame of ``fiducial'' observers, but in the laboratory frame $S$ as
well. Using the LT of $E^{\mu }$ (\ref{f0}) we get
\begin{equation}
K_{L}^{0}=q\beta \gamma E^{\prime 1},K_{L}^{1}=q\gamma E^{\prime
1},K_{L}^{2}=qE^{\prime 2},K_{L}^{3}=qE^{\prime 3}=0.  \label{k}
\end{equation}
Then it can be seen that only $N^{02}=x^{0}K_{L}^{2}-x^{2}K_{L}^{0}$ and $%
N^{12}=x^{1}K_{L}^{2}-x^{2}K_{L}^{1}$ remain. However they are also zero $%
N^{02}=q\beta \gamma (x^{\prime 1}E^{\prime 2}-x^{\prime 2}E^{\prime 1})=0$
and $N^{12}=q\gamma (x^{\prime 1}E^{\prime 2}-x^{\prime 2}E^{\prime 1})=0$,
since $E^{\prime 1,2}=Dx^{\prime 1,2}$. Once again it is obtained that $%
N=(1/2)N^{\mu \nu }\gamma _{\mu }\wedge \gamma _{\nu }=0$.

Of course if instead of using the passive LT of $E$ and $B$ (corresponding
to (\ref{nle})) we deal with the ST (\ref{A1}), i.e., (\ref{ce}), then the
3D magnetic field will appear in the Lorentz force in the $S$ frame. This
will cause that in $S$ the 3D torque will be different from zero and the
principle of relativity will be violated.\bigskip \medskip

\noindent \textit{8.4. The proof that }$N^{\mu \nu }=0$ \textit{in}\textbf{\
}$S$ \textit{using the expressions}\textbf{\ }(\ref{ec})\textbf{\ }\textit{%
for }$E$

\textit{and} $B$. \textit{The frame of ``fiducial'' observers is the}\textbf{%
\ }$S$ \textit{frame }$\bigskip $

\noindent Let us now assume that the laboratory frame $S$ is the frame of
``fiducial'' observers ($v=c\gamma _{0}$, $v^{\mu }=(c,0,0,0)$ in $S$) in
which the temporal components of the 4D $E$ and $B$ are zero and only their
spatial components remain. In the laboratory frame $S$ both charges $Q$ and $%
q$ are moving and the components in the CBGQs $u_{Q}^{\mu }\gamma _{\mu }$
and $u^{\mu }\gamma _{\mu }$ are given as
\begin{equation}
u_{Q}^{\mu }=u^{\mu }=(\gamma c,\gamma \beta c,0,0).  \label{br}
\end{equation}
The fields $E$ and $B$ as AQs are given by (\ref{ec}) and when they are
written as CBGQs in $S$ then $v=c\gamma _{0}$ and the components of $u_{Q}$
are determined by (\ref{br}). The components $E^{\mu }$ become $%
E^{0}=E^{3}=0,$ $E^{1}=D\gamma (x^{1}-\beta x^{0}),$ $E^{2}=D\gamma x^{2}$.
Taking into account that in $S^{\prime }$ $t^{\prime }=0$, i.e., $x^{\prime
0}=\gamma (x^{0}-\beta x^{1})=0$, the relation $x^{0}=\beta x^{1}$ is
obtained. Inserting this last relations into expressions for $E^{\mu }$ we
find
\begin{equation}
E^{0}=E^{3}=0,\ E^{1}=Dx^{1}/\gamma ,\ E^{2}=D\gamma x^{2}.  \label{le}
\end{equation}
The charge\textbf{\ }$Q$ moves in the $S$ frame (now it is the frame of
``fiducial'' observers), which yields that the magnetic field $B=B^{\mu
}\gamma _{\mu }$ is now different from zero. The components $B^{\mu }$ are
\begin{equation}
B^{0}=B^{1}=B^{2}=0,\ B^{3}=(1/c)D\gamma \beta x^{2}=\beta E^{2}/c.
\label{mg}
\end{equation}
The spatial components $E^{i}$ and $B^{i}$ from (\ref{le}) and (\ref{mg})
are the same as the usual expressions for the components of the 3D vectors $%
\mathbf{E}$ and $\mathbf{B}$. Inserting (\ref{le}) and (\ref{mg}) into (\ref
{lo}) we find the expression for the Lorentz force $K_{L}$ in the laboratory
frame $S.$ The components of $K_{L}$ in $S$ are
\begin{eqnarray}
K_{L\parallel }^{0} &=&q\gamma \beta E^{1},\ K_{L\parallel }^{i}=0,\
K_{L\perp }^{0}=K_{L\perp }^{3}=0,  \notag \\
K_{L\perp }^{1} &=&q\gamma E^{1},\ K_{L\perp }^{2}=q\gamma (E^{2}-\beta
cB^{3})=qE^{2}/\gamma .  \label{si}
\end{eqnarray}
We see that in the laboratory frame $S$, when it is the frame of
``fiducial'' observers, there is the 4D magnetic field (\ref{mg}) which
enters into the expression for the total 4D Lorentz force $K_{L}$. Then
using (\ref{le}), (\ref{mg}), (\ref{si}) and the relation $x^{0}=\beta x^{1}$
one easily finds all components $N^{\mu \nu }$
\begin{eqnarray}
x^{3} &=&0,\ K_{L}^{3}=0\Rightarrow N^{03}=N^{13}=N^{23}=0,  \notag \\
K_{L}^{0} &=&\beta K_{L}^{1}\Rightarrow N^{01}=x^{1}(\beta
K_{L}^{1}-K_{L}^{0})=0,  \label{osr} \\
K_{L}^{2} &=&q\gamma (E^{2}-\beta cB^{3})=qE^{2}/\gamma \Rightarrow
N^{02}=N^{12}=0.  \notag
\end{eqnarray}
Thus although in $S$ there is the 4D magnetic field (\ref{mg}) and a part of
$K_{L}$ (in $K_{L}^{2}$ in (\ref{si})), which corresponds to the magnetic
force, it is again obtained that all components $N^{\mu \nu }$ are zero, $%
N^{\mu \nu }=0$, and consequently $N=(1/2)N^{\mu \nu }\gamma _{\mu }\wedge
\gamma _{\nu }=0$. This proof is very instructive since it nicely clarifies
the fundamental difference between the usual approaches with 3D quantities
and our approach with 4D geometric quantities. In the usual approaches the
3D magnetic field $\mathbf{B}$ (which arises from the ST (\ref{Aeb}) of the
3D $\mathbf{E}^{\prime }$) yields the 3D magnetic force $q\mathbf{V}\times
\mathbf{B}$ and this causes that the 3D torque $\mathbf{N}$ is different
from zero in the laboratory frame $S$. On the other hand when geometric 4D
quantities are used then the 4D torque $N$ is zero despite of the fact that
in $S$ the charge $Q$, which is moving in $S$, produces both the 4D $E$ and $%
B$ (given by equation (\ref{ec})). The conclusion that can be drawn from
this proof is that the real cause of the violation of the principle of
relativity and of Jackson's paradox is the use of 3D quantities as physical
quantities in the 4D spacetime.

We see that always the same result (\ref{enc}) is obtained. This
consideration explicitly shows the consistency of the approach with
geometric 4D quantities. In addition the proofs from sections 8.1-8.3 once
again reveal that the relativistically correct transformations of the 4D
electric and magnetic fields, which are in a complete agreement with the
principle of relativity, are the LT (\ref{nle}) and (\ref{nlb}) and not, as
generally believed, the ST (\ref{ce}), (\ref{B}) and (\ref{sk1}) or (\ref
{Aeb}) and (\ref{A1}). \bigskip \medskip

\noindent \textbf{9.}\textit{\ }\textbf{Comparison with the Trouton-Noble
experiment}\bigskip

\noindent The main difference between our geometric approach to the
considered problem and the approach in Jackson's paper [1] is that in the
geometric approach \emph{the independent physical reality is attributed only
to the geometric 4D quantities, AQs or CBGQs, and not, as usual, to the 3D
quantities. }In [1] even the covariant quantities, e.g., $M^{\mu \nu }$, $%
x^{\mu }$, $u^{\nu }$, $F^{\alpha \beta }$, etc. are considered as auxiliary
mathematical quantities from which ``physical'' 3D quantities are deduced.
However the considerations in the preceding sections and in [12-19] show
that the geometric approach is, as already said, in a complete agreement
with the principle of relativity and with experiments, see [14] and [16-18].

In this section we shall discuss the Trouton-Noble experiment [27], see also
[28], comparing the usual explanations with our geometric approach that
explicitly uses AQs or CBGQs. In the experiment they looked for the turning
motion of a charged parallel plate capacitor suspended at rest in the frame
of the earth in order to measure the earth's motion through the ether. The
explanations, which are given until now (see, e.g., [29-33] and references
therein) for the null result of the experiments [27] ([28]) are not correct
from the invariant SR viewpoint, since they use quantities and
transformations that are not well-defined in the 4D spacetime; e.g., the
Lorentz contraction, the nonelectromagnetic forces of undefined nature, the
ST for the 3D $\mathbf{E}$ and $\mathbf{B}$ (\ref{Aeb}) (or (\ref{ce}), (\ref
{B}) and (\ref{sk1})) and for the 3D torque, etc.. In all previous
treatments it is found that there is no 3D torque $\mathbf{N}$ for the
stationary capacitor since there is only a 3D electric force $q\mathbf{E}$
in the rest frame of the capacitor. However, \emph{a 3D torque is always
obtained for the moving capacitor} that is caused by the 3D magnetic force $q%
\mathbf{V}\times \mathbf{B}$; the existence of a 3D magnetic field $\mathbf{B%
}$ in that frame comes from the transformations (\ref{Aeb}). Everything
happens in the same way as in the above discussed Jackson's paradox. Then,
in order to get the agreement with experiments (and with the principle of
relativity), different explanations are offered for the existence of \emph{%
another 3D torque,} which is equal in magnitude but of opposite direction
giving that the total 3D torque is zero. In our approach the explanation for
the null result is very simple and natural; all quantities are invariant 4D
quantities, which means that their values are the same in the rest frame of
the capacitor and in the moving frame. Thus if there is no torque (but now
as a geometric, invariant, 4D quantity) in the rest frame then the capacitor
cannot appear to be rotating in a uniformly moving frame. This explanation
is the same as that one for Jackson's paradox which is presented in sections
8-8.4.

We shall not discuss previous explanations given in [29-32] but only the
recent ``resolution'' presented in [33]. It is argued there that the
Trouton-Noble paradox is resolved once the electromagnetic momentum of the
moving capacitor is properly taken into account. First it is obtained that
there is a 3D mechanical torque on the moving capacitor, due to the 3D
magnetic force, and then it is shown that the rate of change of the 3D
angular electromagnetic field momentum associated with the moving capacitor
completely balances the 3D mechanical torque.The consideration with 4D
quantities and their LT will reveal that there is no need either for the
nonelectromagnetic forces and their torque, [29-32], or for the angular
electromagnetic field momentum and its rate of change, i.e., its torque
[33]. Therefore we shall examine in more detail the calculation of the 3D
torque that is presented in [33], but we do not need to consider the
calculation of the 3D angular electromagnetic field momentum. (It is worth
noting that the resolution of the the Trouton-Noble paradox using geometric
4D quantities is already presented in [19], but there we dealt with the
electromagnetic field $F$.)\bigskip \medskip

\noindent \textit{9.1. Jefimenko's calculation }[33]\textit{\ of the 3D
torque\bigskip \medskip }

\noindent First let us discuss Jefimenko's calculation [33] of the 3D
torque. The rest frame of a thin parallel-plate capacitor is the $S^{\prime
} $ frame. In the $S$ frame the capacitor moves with uniform 3D velocity $%
\mathbf{V}$ in the positive direction of the $x^{1}$ - axis. (figure 1 from
[33] is actually a projection onto the hypersurface $t^{\prime }=const.$,
which means that $x$, $y$ and $\Theta $ from that figure 1 would need to be
denoted as $x^{\prime 1}$, $x^{\prime 2}$ and $\Theta ^{\prime }$
respectively.) In the $S^{\prime }$ frame $A$ denotes the surface area of
the capacitor's plates, $a$ is the distance between the capacitor's plates
and $\Theta ^{\prime }$ is the angle between the line joining the axis of
rotation (i.e., the middle of the negative plate) with the middle of the
positive plate and the $x^{\prime 2}$ axis. That line is taken to be in the $%
x^{\prime 1}$, $x^{\prime 2}$ plane. The uniform surface charge density on
the capacitor's plates is $\sigma $. In the $S^{\prime }$ frame there is
only a 3D electric force and, in the same way as in Jackson's paradox with
two charges, the 3D torque on the stationary capacitor is zero $\mathbf{N}%
^{\prime }=\mathbf{r}^{\prime }\times \mathbf{F}^{\prime }=0$; the total 3D
force $\mathbf{F}^{\prime }$ and $\mathbf{r}^{\prime }$ are along the same
line. In components $\mathbf{N}^{\prime }$ becomes
\begin{equation}
N_{i}^{\prime }=\varepsilon _{ijk}r_{j}^{\prime }F_{k}^{\prime }=0,
\label{in}
\end{equation}
where $F_{k}^{\prime }$ are the components of the total 3D force (electric)
acting on the positive plate of the stationary capacitor, $F_{1}^{\prime
}=(\sigma ^{2}A/2\varepsilon _{0})\sin \Theta ^{\prime }$, $F_{2}^{\prime
}=-(\sigma ^{2}A/2\varepsilon _{0})\cos \Theta ^{\prime }$ and $%
F_{3}^{\prime }=0$, and $r_{i}^{\prime }$ are the components of the lever
arm joining the axis of rotation with the point of application of the
resultant 3D force, i.e., the midpoint of the positive plate,
\begin{equation}
r_{1}^{\prime }=-a\sin \Theta ^{\prime },r_{2}^{\prime }=a\cos \Theta
^{\prime },r_{3}^{\prime }=0,  \label{ri}
\end{equation}
see figure 1 in [33]. (The 3D electric field produced by the negative plate
of the capacitor at the location of the positive plate is
\begin{equation}
\mathbf{E}_{-}^{\prime }=(-\sigma /2\varepsilon _{0}a)(r_{1}^{\prime }%
\mathbf{i}^{\prime }+r_{2}^{\prime }\mathbf{j}^{\prime }),  \label{mi}
\end{equation}
whence $\mathbf{F}^{\prime }=q\mathbf{E}_{-}^{\prime }=(\sigma A)\mathbf{E}%
_{-}^{\prime }$, or in components
\begin{equation}
F_{1}^{\prime }=Cr_{1}^{\prime },F_{2}^{\prime }=Cr_{2}^{\prime
},F_{3}^{\prime }=0,\ C=-\sigma ^{2}A/2\varepsilon _{0}a.  \label{f5}
\end{equation}
$\mathbf{i}^{\prime }$ and $\mathbf{j}^{\prime }$ are unit 3D vectors in the
direction of the $x^{\prime }$ - and $y^{\prime }$ - axis, respectively and $%
q$ is the total charge residing on the positive plate.) In equations (\ref
{in})-(\ref{f5}) the components of the 3D vectors $\mathbf{N}^{\prime }$, $%
\mathbf{r}^{\prime }$ and $\mathbf{F}^{\prime }$ are written with lowered
(generic) subscripts, since they are not the spatial components of 4D
quantities. This refers to the third-rank antisymmetric $\varepsilon $
tensor too. The super- and subscripts are used only on components of 4D
quantities. Then in [33] the 3D torque experienced by the moving capacitor
is determined by using ``relativistic'' (my quotation-marks) transformation
equations for the torque. These ``relativistic'' transformation equations
for the 3D torque given in [33] are
\begin{equation}
N_{1}=N_{1}^{\prime }/\gamma ,\ N_{2}=N_{2}^{\prime }+\beta
^{2}r_{1}^{\prime }F_{3}^{\prime },\ N_{3}=N_{3}^{\prime }-\beta
^{2}r_{1}^{\prime }F_{2}^{\prime },  \label{s}
\end{equation}
where $\beta =V/c$, $\gamma =(1-\beta ^{2})^{-1/2}$. Equations (\ref{s}) are
equations (1)-(3) in [33].

The transformations (\ref{s}) of the 3D $\mathbf{N}$ are found, e.g., in
Jefimenko's book [34]. In section 8 in [34], under the title: ``From
relativistic electromagnetism to relativistic mechanics,'' the
transformations of different 3D quantities are presented. Among others in
section 8-6 in [34] the transformations of a 3D torque are presented.
Jefimenko [34], as all others, considers that the transformations (\ref{s})
are the LT, but we shall call them the ST of the 3D $\mathbf{N}$ (in analogy
with the ST of the 3D $\mathbf{E}$ and $\mathbf{B}$ (\ref{Aeb}) or (\ref{A1}%
)) since they are not the LT of 4D quantities. The same name, the ST, will
be used for the transformations of all other 3D quantities, e.g., the usual
transformations of components of the 3D angular momentum $\mathbf{L}$ that
are given by equation (11) in [1], then the transformations of the 3D force $%
\mathbf{F}$ that are given by equations (8-5.1)-(8-5.3) in [34], or by
equations (1.56)-(1.58) in [35], then the well-known transformations of the
3D velocity $\mathbf{V}$ given, e.g., by equations (11.31) in [2], or
equations (7-2.5)-(7-2.7) in [34], etc. All mentioned transformations of the
3D quantities are, in fact, the ``apparent'' transformations that are
discussed in section 6.

Now let us proceed with the derivation of the 3D torque from [33]. Taking
into account in the ST of the 3D $\mathbf{N}$ (\ref{s}) that $N_{i}^{\prime
}=0$ (\ref{in}) and $F_{3}^{\prime }=0$ Jefimenko [33] finds that $N_{3}$
component is different from zero
\begin{equation}
N_{3}=-(V^{2}/c^{2})r_{1}^{\prime }F_{2}^{\prime }.  \label{k5}
\end{equation}
This result is commented in [33] in the following way: \emph{``We have thus
obtained a paradoxical result: contrary to the relativity principle,
although our stationary capacitor experiences no torque, the same capacitor
moving with uniform velocity along a straight line appears to experience a
torque. What makes this result especially surprising is that we have arrived
at it by using relativistic transformations that are based on the very same
relativity principle with which they now appear to conflict.''} (my
emphasis) Thus again the same paradox arises with the violation of the
principle of relativity as in the above discussed Jackson's paradox. It is
assumed in [33], as in many other papers including [29-32], that the
transformations (\ref{s})) are the relativistic transformations, i.e., the
LT, that are based on the principle of relativity. Such opinion implicitly
supposes that 3D quantities, their transformations and physical laws written
in terms of them are physically real in the 4D spacetime and in agreement
with the principle of relativity. Actually such opinion prevails already
from Einstein's fundamental work on SR [4].\bigskip \medskip

\noindent \textit{9.2. Resolution of the Trouton-Noble paradox in the
invariant SR\bigskip \medskip }

\noindent The approach of the invariant SR [12-19] is completely different.
There, as already explained, the physical reality in the 4D spacetime is
attributed only to geometric 4D quantities, AQs or CBGQs, their LT and
physical laws written in terms of them. The principle of relativity is
automatically included in such formulation.

Thence in the 4D spacetime we are dealing with the Lorentz force $%
K=(q/c)F\cdot u$, where $u$ is the velocity 1-vector of a charge $q$. The
torque $N$, as a 4D AQ, is defined as a bivector
\begin{equation}
N=r\wedge K,\ r=x_{P}-x_{O},  \label{or}
\end{equation}
where $r$ is 1-vector associated with the lever arm, $x_{P}$ and $x_{O}$ are
the position 1-vectors associated with the spatial point of the axis of
rotation and the spatial point of application of the force $K$, $P$ and $O$
are the events whose position 1-vectors are $x_{P}$ and $x_{O}$.

In general, as in (\ref{MKN}), the proper velocity $u$ for a point particle
is $u=dx/d\tau $, $\tau $ is the proper time, $p$ is the proper momentum $%
p=mu$, the proper angular momentum of a particle is the bivector $M=x\wedge p
$ and the torque $N$ about the origin is the bivector $N=dM/d\tau =x\wedge K$%
, where in this relation $K$ is an arbitrary force 1-vector. When $K$ is
written as a CBGQ in the standard basis $\{\gamma _{\mu }\}$ then its
components are $K^{\mu }=(\gamma _{u}F_{i}U_{i}/c,\gamma _{u}F_{1},\gamma
_{u}F_{2},\gamma _{u}F_{3})$, and the components of $u$ in the $\{\gamma
_{\mu }\}$ basis are $u^{\mu }=(\gamma _{u}c,\gamma _{u}U_{1},\gamma
_{u}U_{2},\gamma _{u}U_{3})$. $\gamma _{u}=(1-U^{2}/c^{2})^{-1/2}$, $F_{i}$
are components of the 3D force $\mathbf{F}$ and $U_{i}$ are components of
the 3D velocity $\mathbf{U}$. We see that only when the considered particle
is at rest, i.e., $U_{i}=0$, $\gamma _{u}=1$ and consequently $u^{\mu
}=(c,0,0,0)$, then $K^{\mu }$ contains only the components $F_{i}$, i.e., $%
K^{\mu }=(0,F_{1},F_{2},F_{3})$. However even in that case $u^{\mu }$ and $%
K^{\mu }$ are the components of geometric \emph{4D} \emph{quantities }$u$
and $K$ in the $\{\gamma _{\mu }\}$ basis and not the components of some
\emph{3D quantities} $\mathbf{U}$ and $\mathbf{F}$. The LT correctly
transform the whole 4D quantity, which means that there is no physical sense
in such transformations like (\ref{s}); these transformations are not
relativistic and they are not based on the principle of relativity. All
conclusions derived from such relations as are equations (\ref{s}) have
nothing in common with SR as the theory of the 4D spacetime.

After this digression we go back to the resolution of the Trouton-Noble
paradox in the invariant SR. Since we have the same problem as in the above
discussed Jackson's paradox we could use any of the proofs from sections 8.1
- 8.4, but, for simplicity, we shall consider only the proof from section
8.1. As in section 8.1 the $S^{\prime }$ frame is the frame of ``fiducial''
observers and it is the rest frame of the capacitor. In that frame we choose
that $r^{\prime 0}=x_{P}^{\prime 0}-x_{O}^{\prime 0}=0$. The system of
coordinates is chosen in such a way that $r^{\prime 3}=0$ (as in figure 1 in
[33]) giving that $r^{\prime }=r^{\prime \mu }\gamma _{\mu }^{\prime }$, $%
r^{\prime \mu }=(0,r_{1}^{\prime },r_{2}^{\prime },0)$, where $r^{\prime i}$
are the same as in (\ref{ri}) (remember the convention about lowered
(generic) subscripts for 3D quantities that is mentioned in connection with
equation (\ref{in})). Further, for the same reasons as in section 8.1, we
have that $K^{\prime \mu }=(0,F_{1}^{\prime },F_{2}^{\prime },0)$, where $%
K^{\prime 1}$ and $K^{\prime 2}$ are the same as in (\ref{f5}), i.e.,
\begin{equation}
K^{\prime }=(\sigma A)(E^{\prime 1}\gamma _{1}^{\prime }+E^{\prime 2}\gamma
_{2}^{\prime })=C(r^{\prime 1}\gamma _{1}^{\prime }+r^{\prime 2}\gamma
_{2}^{\prime }).  \label{k0}
\end{equation}
This yields that $N^{\prime i0}=N^{\prime 13}=N^{\prime 23}=0$ and only
remains $N^{\prime 12}=r^{\prime 1}K^{\prime 2}-r^{\prime 2}K^{\prime 1}$,
which, taking into account (\ref{k0}), becomes
\begin{equation}
N^{\prime 12}=C(r^{\prime 1}r^{\prime 2}-r^{\prime 2}r^{\prime 1})=0.
\label{e7}
\end{equation}
Thus all $N^{\prime \alpha \beta }$ are zero in the $S^{\prime }$ frame in
which the capacitor is at rest. Since the CBGQ $(1/2)N^{\prime \mu \nu
}\gamma _{\mu }^{\prime }\wedge \gamma _{\nu }^{\prime }$ is an invariant
quantity upon the passive LT we have proved, as in section 8.1, that not
only the components $N^{\prime \alpha \beta }$ are zero but at the same time
that the whole torque $N$ is zero
\begin{equation}
N=(1/2)N^{\prime \mu \nu }\gamma _{\mu }^{\prime }\wedge \gamma _{\nu
}^{\prime }=(1/2)N^{\mu \nu }\gamma _{\mu }\wedge \gamma _{\nu }=0.
\label{en1}
\end{equation}
Thence the torque is zero not only for the stationary capacitor but for the
moving capacitor as well. We see that in the approach with geometric 4D
quantities there is no Trouton-Noble paradox, as there is no Jackson's
paradox. \bigskip \medskip

\noindent \textbf{10. Conclusions \bigskip }

\noindent In both considered paradoxes there is a 3D torque and so a time
rate of change of 3D angular momentum in one inertial frame, but no 3D
angular momentum and no 3D torque in another. The principle of relativity is
violated and also there is no agreement with the Trouton-Noble experiment.
In all usual approaches the 3D magnetic field $\mathbf{B}$ arises from the
ST (\ref{Aeb}) of the 3D $\mathbf{E}^{\prime }$. This $\mathbf{B}$ field
determines the 3D magnetic force $q\mathbf{V}\times \mathbf{B}$ and this
causes that the 3D torque $\mathbf{N}$ is different from zero in the
laboratory frame $S$, where the charges are moving. However the proofs from
sections 8.1-8.4 and 9.2 reveal that the relativistically correct
transformations of the 4D electric and magnetic fields, which are in a
complete agreement with the principle of relativity, are the LT (\ref{nle})
and (\ref{nlb}) and not, as generally believed, the ST (\ref{ce}), (\ref{B})
and (\ref{sk1}) or (\ref{Aeb}) and (\ref{A1}). In our geometric approach,
i.e., in the invariant SR [12-19], the independent physical reality in the
4D spacetime is attributed only to geometric 4D quantities, AQs or CBGQs,
their LT and physical laws written in terms of them and not, as usual, to
the 3D quantities. When geometric 4D quantities are used then it is
consistently obtained in different manners that the 4D torque $N$ is always
zero, see sections 8.1-8.4 and 9.2. The principle of relativity is
automatically satisfied with such quantities and there is not either
Jackson's paradox or the Trouton-Noble paradox. The main conclusion that can
be drawn from the whole consideration in this paper is that the
relativistically correct description of physical phenomena can be achieved
with geometric 4D quantities as physical quantities in the 4D spacetime and
not, as usual, with 3D quantities. This conclusion is in a full agreement
with all other results obtained in [12-19] and [23]. \bigskip \medskip

\noindent \textbf{References\bigskip }

\noindent \lbrack 1] Jackson J D 2004 \textit{Am. J. Phys.} \textbf{72} 1484

\noindent \lbrack 2] Jackson J D 1977 \textit{Classical Electrodynamics}
(New York: Wiley) 2nd

edn.

\noindent \lbrack 3] Lorentz H A 1904 \textit{Proceedings of the Academy of
Sciences of Amsterdam}

6 in Perrett W and Jeffery G B 1952 \textit{The Principle of Relativity }%
(New York:

Dover)

\noindent \lbrack 4] Einstein A 1905 \textit{Ann. Physik.} \textbf{17} 891
tr. by Perrett W and Jeffery

G B 1952 in \textit{The Principle of Relativity} (New York: Dover)

\noindent \lbrack 5] Poincar\'{e} H 1906 \textit{Rend. del Circ. Mat. di
Palermo} \textbf{21} 129

\noindent \lbrack 6] Logunov A A 1996 \textit{Hadronic J.} \textbf{19} 109

\noindent \lbrack 7] Hestenes D 1966 \textit{Space-Time Algebra }(New York:
Gordon and Breach)

\noindent \lbrack 8] Hestenes D 1999 \textit{New Foundations for Classical
Mechanics }(Dordrecht:

Kluwer) 2nd. edn.

\noindent \lbrack 9] Hestenes D 2003 \textit{Am. J Phys.} \textbf{71 }691

\noindent \lbrack 10] Doran C and Lasenby A 2003 \textit{Geometric algebra
for physicists }(Cambridge:

Cambridge University Press)

\noindent \lbrack 11] Jancewicz B 1989 \textit{Multivectors and Clifford
Algebra in Electrodynamics}

(Singapore: World Scientific)

\noindent \lbrack 12] Ivezi\'{c} T 1999 \textit{Found. Phys. Lett.} \textbf{%
12 }507

\noindent \lbrack 13] Ivezi\'{c} T 2001 \textit{Found. Phys.} \textbf{31}
1139

\noindent \lbrack 14] Ivezi\'{c} T 2002 \textit{Found. Phys. Lett.} \textbf{%
15} 27; \textit{physics}/0103026;

\textit{physics}/0101091.

\noindent \lbrack 15] Ivezi\'{c} T \textit{hep-th}/0207250v2.

\noindent \lbrack 16] Ivezi\'{c} T 2003 \textit{Found. Phys.} \textbf{33}
1339

\noindent \lbrack 17] Ivezi\'{c} T \textit{physics}/0411166 to be published
in 2005 \textit{Found. Phys. Lett.}

\noindent \lbrack 18] Ivezi\'{c} T \textit{physics}/0409118 to be published
in 2005 \textit{Found. Phys.}

\noindent \lbrack 19] Ivezi\'{c} T \textit{physics}/0412167

\noindent \lbrack 20] Wald R M 1984 \textit{General Relativity} (Chicago:
The University of

Chicago Press); Ludvigsen M 1999 \textit{General Relativity,} \textit{A
Geometric }

\textit{Approach }(Cambridge: Cambridge University Press); Vanzella

D A T Matsas G E A and Crater H W 1996 \textit{Am. J. Phys.} \textbf{64}
1075;

Sonego S and Abramowicz M A 1998 \textit{J. Math. Phys.} \textbf{39} 3158

\noindent \lbrack 21] N\'{u}\~{n}ez Y\'{e}pez H N Salas Brito A L and Vargas
C A 1988 \textit{Revista}

\textit{Mexicana de F\'{i}sica} \textbf{34} 636

\noindent \lbrack 22] Esposito S 1998 \textit{Found. Phys.} \textbf{28} 231

\noindent \lbrack 23] Ivezi\'{c} T 1999 \textit{Found. Phys. Lett.} \textbf{%
12} 105

\noindent \lbrack 24] Hestenes D and Sobczyk G 1984 \textit{Clifford Algebra
to Geometric Calculus}

(Dordrecht: Reidel).

\noindent \lbrack 25] Rohrlich F 1966 \textit{Nuovo Cimento B} \textbf{45} 76

\noindent \lbrack 26] Gamba A 1967 \textit{Am. J. Phys.} \textbf{35} 83

\noindent \lbrack 27] Trouton F T and Noble H R 1903 \textit{Philos. Trans.
R. Soc. London }

\textit{Ser. A }\textbf{202} 165

\noindent \lbrack 28] Hayden H C 1994 \textit{Rev. Sci. Instrum.} \textbf{65}
788

\noindent \lbrack 29] Singal A K 1993 \textit{Am. J. Phys.} \textbf{61} 428

\noindent \lbrack 30] von Laue M 1911 \textit{Phys. Zeits.} \textbf{12} 1008

\noindent \lbrack 31] Pauli W 1958 \textit{Theory of Relativity} (New York:
Pergamon)

\noindent \lbrack 32] Teukolsky S A 1996 \textit{Am. J. Phys.} \textbf{64}
1104

\noindent \lbrack 33]Jefimenko O D 1999 \textit{J. Phys. A: Math. Gen.}
\textbf{32} 3755

\noindent \lbrack 34] Jefimenko O D 1997 \textit{Retardation and Relativity}
(Star City: Electret

Scientific)

\noindent \lbrack 35] Rosser W G W 1968 \textit{Classical Electromagnetism
via Relativity}

(New York: Plenum Press)

\end{document}